\documentclass{aa}
\usepackage{float}
\usepackage{array}
\usepackage{tabularx}
\usepackage{changepage}
\usepackage{color}
\usepackage[latin1]{inputenc} 
\usepackage{graphicx}
\usepackage{natbib}
\usepackage{mathabx}
\usepackage{multirow}
\usepackage{txfonts}
\usepackage{bm}

\begin{document}
\title{Magnetic field structure and halo in NGC~4631 \thanks{Based on observations
with the 100 m telescope of the MPIfR (Max-Planck-Institut f\"ur Radioastronomie) at Effelsberg and the VLA operated by the NRAO. The NRAO is a facility of the National Science Foundation operated under agreement by Associated Universities, Inc.}}
\author{Silvia Carolina Mora \thanks{\email{caromora@mpifr-bonn.mpg.de}}
\and Marita Krause \thanks{\email{mkrause@mpifr-bonn.mpg.de}}}
\institute{Max-Planck-Institut f\"ur Radioastronomie, Auf dem H\"ugel 69, 53121 Bonn, Germany}
\date{Received 4 January 2013 / Accepted 24 September 2013}

\abstract{All of the edge-on spiral galaxies observed so far present a similar magnetic field configuration, which consists of a plane-parallel field in the disk and an X-shaped field at larger z-distances from the plane of the galaxy. Only NGC~4631 seems to have a different field orientation in its disk. Along the eastern and western halves of the disk of NGC~4631 the magnetic field orientation is parallel to the galactic plane, but in the central region of the disk a vertical field seems to dominate.} 
{In order to clarify whether NGC~4631 has a unique magnetic field configuration in the central region along its disk, we present high-resolution Faraday-corrected polarization data.}
{Radio continuum observations of NGC~4631 at 4.85~GHz were performed with the VLA. In addition, observations were made with the Effelsberg telescope at 4.85~GHz and at 8.35~GHz. These were analyzed together with archival VLA-data at 8.35~GHz. The single-dish and interferometer data were combined to recover the missing zero-spacings.}
{We determined an integrated total spectral index of $\alpha_{tot} = -0.78 \pm 0.04$ and a nonthermal integrated spectral index of $\alpha_{nth} = -0.87 \pm 0.03$. The vertical scale heights in NGC~4631 vary significantly in different regions within the galaxy and their mean values at 4.85~GHz are with 2.3~kpc (370~pc) for the thick (thin) disk higher than the mean values found so far in six other edge-on spiral galaxies. This may originate in the tidal interaction of NGC~4631 with its neighbouring galaxies. The rotation measures are characterized by a smooth large-scale distribution. Along the galactic plane the degree of Faraday depolarization is significantly high. We estimated a total magnetic field strength in the disk of NGC~4631 of $B_{ t}\approx 9\pm2~\rm{\mu G}$ and an ordered field strength of $B_{ ord} \approx  2 \pm 1~\rm{ \mu  G}$. The total field strengths in the halo are of the order of the total magnetic field strength in the disk, whereas the ordered field strengths in the halo seem to be higher than the value in the disk.}
{The derived distribution of rotation measures implies that NGC~4631 has a large-scale regular magnetic field configuration. Despite the strong Faraday depolarization along the galactic plane and the strong beam depolarization in the transition zone between the disk and halo, our research strongly indicates that the magnetic field orientation along the central 5-7~kpc of the disk is also plane-parallel. Therefore, we claim that NGC~4631 also has a magnetic field structure plane-parallel along its entire disk, similar to all other edge-on galaxies observed up to now.}

\maketitle

\section{Introduction}
\label{intro}

In external galaxies the best tracer of the magnetic field is synchrotron radiation produced by relativistic electrons moving in a magnetic field. These particles emit electromagnetic waves in a characteristic nonthermal frequency spectrum and with a polarization direction perpendicular to the magnetic field. The plane of polarization of the electromagnetic waves rotates as the linearly polarized emission passes through a magnetized plasma along the line of sight (LOS). This phenomenon is known as the Faraday effect and the angle by which it rotates ($\Delta \psi$) in a Faraday thin medium is proportional to the rotation measure (RM):  $\Delta \psi =\lambda^2\cdot RM$. Furthermore, the RM is proportional to the line-of-sight integral over the density of thermal electrons multiplied by the strength of the regular field component parallel to the LOS: $RM~\propto~\int n_{e}B_\shortparallel dl$ \citep{Gardner}.

With the RM distribution one can correct the polarization angles for Faraday rotation and also obtain information about the strength (if we have knowledge of n$_e$) and the direction of the magnetic field component along the LOS: the sign of the rotation measures indicates the direction of the B$_\shortparallel$-field. By definition, positive values indicate that the B$_\shortparallel$-field is directed towards the observer and negative values indicate that the field is directed away from the observer. In addition, rotation measure values add up for parallel fields along the LOS, but may cancel for antiparallel fields of equal strength.

The best way to study magnetic fields in galaxies is with radio observations of the continuum emission in the cm-wavelength regime. The total intensity of the synchrotron emission measures the strength of the total magnetic field, while the linearly polarized intensity reveals the strength and the structure of the ordered field perpendicular to the LOS.

Observations of face-on spiral galaxies show that the magnetic field lines in the disk follow a spiral structure similar to the optical spiral morphology \citep{Models2}. Where spiral arms are visible, the fields tend to be more ordered in the interarm regions.
In all of the edge-on spiral galaxies studied up to now, the large-scale ordered fields in the disk lie plane-parallel along the disk of the galaxy and in the halo they have an X-shaped morphology \citep{Krause2009}, sometimes with nearly vertical field components above and below the central region (e.g., in NGC~5775, \citealt{Soida2011}). The disk-parallel magnetic field is the expected edge-on projection of the spiral magnetic disk field seen in face-on galaxies. This large-scale magnetic field in the disk is believed to be amplified by the action of a large-scale $\alpha\Omega$~dynamo \citep{Magnetism}.

The galaxy NGC~4631 is known for the strong vertical magnetic fields in its radio halo \citep{Hummel+Beck}. It has been argued for quite some time that NGC~4631 seems to be the only edge-on galaxy that does not exhibit a plane-parallel field in its disk. However, the large-scale vertical field in the disk in NGC~4631 seems to be restricted to the central region \citep{Krause2003}. Outside of a radius of about 2.5~kpc the field is plane-parallel in the western and eastern halves of the disk. In its outer halo it does present the typical X-shaped configuration seen in the other edge-on galaxies \citep{Golla,Krause2009}. Furthermore, above and below the galactic center the halo field is orientated perpendicular to the plane of the galaxy.

In addition, NGC~4631 may be undergoing considerable star formation activity; this is indicated by the high star formation efficiency \citep{Krause2011}, the blue color (B-V), the strong H$\alpha$ emission \citep{Gollaalone}, and the small ratio $S_{100\rm{\mu m}}/S_{60\rm{\mu m}}$ \citep{Hummel+Beck} of this galaxy. Hence, it was argued that the field in the central region of the disk might be wind-driven and that it may be related to the high star-forming activity in this galaxy \citep{Golla}. However, there are other edge-on galaxies with higher star formation activity that still present plane-parallel magnetic fields in their disks, for example NGC~253 \citep{Krause2011,NGC253}. 

The rotation curve of NGC~4631 rises nearly rigidly in the inner 2.5~\rm{kpc} \citep{Golla+Wielebinski}. Therefore, \cite{Krause2009} has suggested that the differential rotation in this region might be too low to trigger a large-scale $\alpha\Omega$~dynamo and thus that there may be no amplification of the plane-parallel magnetic field in the disk. Nevertheless, it still remains unclear why NGC~4631 may have a unique magnetic field orientation in the central region of its disk. 
However, \cite{CHANGESII} already noticed in their 6~GHz EVLA observations that the magnetic field (uncorrected for Faraday rotation) along the plane of the galaxy appears to be parallel to the disk.

The galaxy NGC~4631 exhibits an extraordinary prominent radio halo \citep{Ekers,Wielebinski+Kap,Hummel+Dettmar}. In addition, the halo has been observed in other interstellar medium components, for example in X-ray emitting gas \citep{Wang,Yamasaki}, dust \citep{Martin, Neininger1999}, and molecular gas \citep{Rand,center}. The central region has an interesting structure consisting of three collinear emission peaks \citep{Ekers,Duric,Gollaalone}. The easternmost radio peak of the triple source coincides with the huge HII region CM~67 \citep{CM67,Roy}. According to \cite{CM67Krause}, the nonthermal radio continuum, CO, and HII line emission in this area are physically connected and form a huge star forming region in NGC~4631.

The interaction of NGC~4631 with its neighbours is argued to be responsible for the large extent and asymmetries of its halo. The galaxy NGC~4631 is located in a group environment where many other galaxies reside \citep{Giuricin}. Among others, there is the dwarf elliptical galaxy NGC~4627 ($2\farcm6$ to the northwest) and the spiral galaxy NGC~4656 (32$'$ to the southeast). The interaction between NGC~4631 and NGC~4656 was discovered by \cite{Roberts}, who mapped the emission from these galaxies in the 21~cm hydrogen radio line. Later on, \cite{Weliachew1978} and \cite{Rand1994} observed these galaxies with the Westerbork Synthesis Radio Telescope and they found an HI bridge emerging from NGC~4631 towards NGC~4656, an HI concentration stretching out to the south of NGC~4631, and two elongated spurs on the northern extension of NGC~4631 (one oriented north-south and the other extending to the north-east), labeled 1 to 4 in the above mentioned papers. \cite{Combes1978} presented a model for the tidal interaction of the three dominant galaxies NGC~4631, NGC~4656, and NGC~4627. Furthermore, gravitational interaction is also known to modify galactic magnetic fields \citep{Drzazga,Chyzy}.

In this paper we present observations of NGC~4631 with the Effelsberg 100 m telescope at 4.75~GHz and 8.35~GHz, and with the VLA at 4.75~GHz. The VLA observations were combined with previously published VLA observations by \cite{Golla} and then merged with the Effelsberg single-dish maps (see Sect.~\ref{Observations}). The results for total intensity, halo  properties, polarized intensity (including Faraday rotation and depolarization), and the magnetic field strengths are presented in Sect.~\ref{Results}, followed by the discussion in Sect.~\ref{Discussion}. The conclusions are summarized in Sect.~\ref{Conclusions}.

The parameters of NGC~4631 assumed throughout this study are presented in Table \ref{NGC}. The dynamical center refers to the location of the central concentration of mass in NGC~4631, which according to \cite{center} is best represented by the IR center at 2~$\rm{\mu m}$.

\begin{table}[h]
\begin{minipage}{1\columnwidth}
\centering
\caption[Parameters of NGC~4631]{\small{Parameters of NGC~4631, as obtained from the literature.}}
\label{NGC}
\begin{tabularx}{1\columnwidth}{c*2{>{\centering\let\\=\tabularnewline}X}}
\hline
\textbf{Parameter} & \textbf{Value}\\ \hline \hline
Morphological type & SBcd \\
\multirow{2}{*}{Dynamical Center}\footnote{IR center at 2~$\rm{\mu m}$ \citep{center}.} & $\alpha_{2000}=12^h42^m08^s$ \\
                                                                                         & $\delta_{2000}= 32^{\circ}32'29 \farcs4$ \\
Inclination & $86^{\circ}$ \\
Position angle of major axis & $85^{\circ}$ \\
Assumed distance\footnote{from \cite{Distance}, determined by the tip of the red giant branch method.} & $7.6~\rm{Mpc}$ ($10'' \corresponds 370~\rm{pc}$) \\ \\ \hline
\end{tabularx}
\end{minipage}
\end{table}

\section{Observations and data reduction}
\label{Observations}

\subsection{Effelsberg observations at 4.85~GHz and 8.35~GHz}

Observations with the 100 m Effelsberg telescope of NGC~4631 at 4.85~GHz ($\lambda$~6.2~cm) were performed in 1996 with the dual-beam receiver (Project 44-95). About 20 coverages for all three Stokes parameters I, U, and Q were made of the $33'\times23'$ scanned area, centered on the dynamical center of the galaxy (see Table \ref{NGC}). The data were reduced with the NOD2 software package \citep{Haslam} in which scanning effects due to weather conditions, receiver instabilities, and the radiation frequency interference (RFIs) were removed. Flux-calibration was done by using the radio source 3C286 on the scales of \cite{Baars}. The U- and Q-maps were used to determine the linearly polarized intensity $I_{pol} = \sqrt{Q^2 + U^2}$ and the polarization angle $ \psi = 1/2 \arctan (U / Q)$. The linear resolution of the maps is $147''$ half-power beam width (HPBW) and the rms noise values are $400~\rm{\mu Jy/beam}$ in total intensity and $73~\rm{\mu Jy/beam}$ in polarized intensity. 

Single dish observations with the Effelsberg 100~m telescope at 8.35~GHz ($\lambda$~3.6~cm) were made with the single-horn receiver at excellent weather conditions between March and May 2002 (Project 134-01) centered on the dynamical center of NGC~4631. Approximately 27 coverages of all three Stokes parameters were taken with a map size of $30'\times20'$ each, scanned in two orthogonal directions to reduce scanning effects. Flux-calibration was again done by using the radio source 3C286 on the scales of \cite{Baars}. High-frequency noise was removed from the maps together with a slight smoothing to $85''$~HPBW in order to increase the signal-to-noise ratio. Again, the polarized intensity and polarization angles were determined from the U- and Q-maps. The rms noise values in the maps are $350~\rm{\mu Jy/beam}$ in total intensity and $70~\rm{\mu Jy/beam}$ in polarized intensity.

\begin{figure}[h]
\centering
\includegraphics[angle=270,trim = 2mm 0mm 0mm 0mm, clip, width=1\columnwidth]{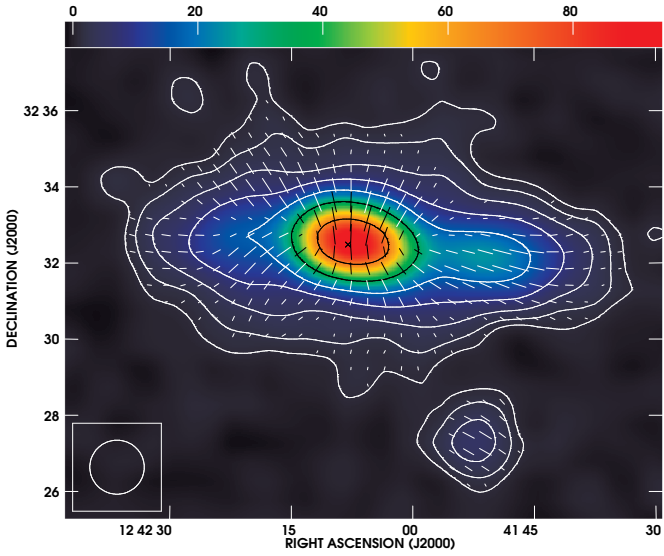}
\caption[]{\small{Effelsberg Stokes I Map at $\lambda$~3.6~cm of NGC~4631 with apparent magnetic field orientation. The length of the vectors is proportional to the polarized intensity $( 1''\corresponds 4 1\; \rm{\mu Jy/beam})$. The colorscale gives the flux density in mJy/beam. The angular resolution is $85''$. The rms noise~($\sigma$) is of $350~\rm{\mu Jy/beam}$. Contour levels correspond to $\sigma\cdot(-3,3,6,12,24,48,96,192,384)$. The ``x'' pinpoints the dynamical center.}}\label{3.6EFFTP}
\end{figure}

\begin{figure}[h]
\centering
\includegraphics[angle=270, trim = 2mm 0mm 0mm 0mm, clip, width=1\columnwidth]{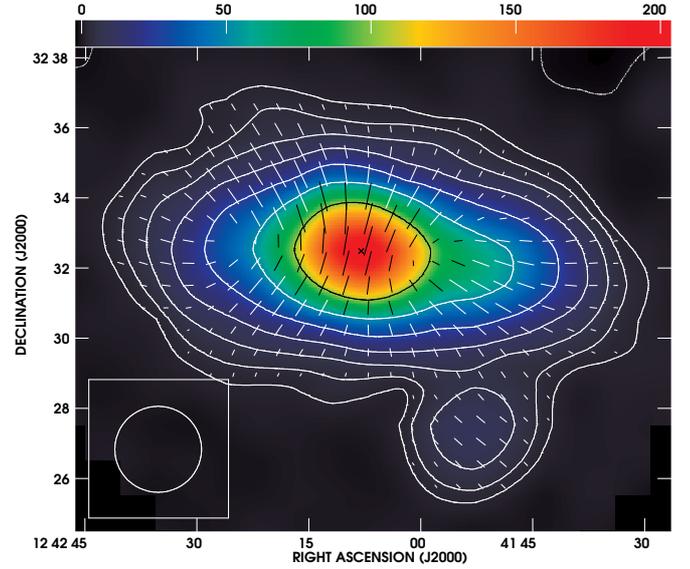}
\caption[]{\small{Effelsberg Stokes I map at $\lambda$~6.2~cm of NGC~4631 with apparent magnetic field orientation. The length of the vectors is proportional to the polarized intensity $(1''\corresponds 56 \;\rm{\mu Jy/beam})$. The colorscale corresponds to the flux density in mJy/beam. The half-power beam width is $147''$ and the rms noise ($\sigma$) is $400~\rm{\mu Jy/beam}$. Contour levels are given by $\sigma\cdot(-3,4,8,16,32,64,128,256,512,1024,2048)$.}}\label{6.2EFF}
\end{figure}

The Effelsberg Stokes I maps together with the linear polarization at $\lambda$~3.6 and 6.2~cm can be seen in Figs. \ref{3.6EFFTP} and \ref{6.2EFF}, respectively. The vectors show the observed E-vectors rotated by $90\degr$, and give the apparent magnetic field orientation since they have not been corrected for Faraday rotation (see Sect.~\ref{RM}). In all of the maps presented in this article the dynamical center of NGC~4631 is indicated by an ``x'' and the beam-area is shown in the left-hand corner of each image.

\subsection{VLA observations at 4.85~GHz (VLA$_{1}$)}

\begin{figure}[h]
\centering
\includegraphics[angle=-90, trim = 2mm 0mm 0mm 0mm, clip,width=1\columnwidth]{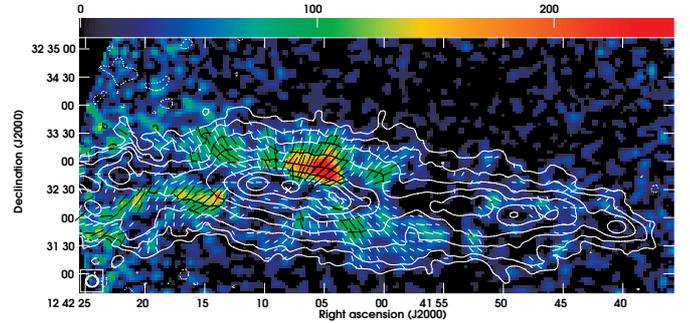}
\caption[]{\small{VLA$_1$ map at $\lambda$~6.2~cm of NGC~4631 with apparent magnetic field orientation. Contours represent the total radio intensity and the colorscale shows the polarized emission in $\rm{\mu Jy/beam}$. The length of the vectors is proportional to the polarized intensity $(1'' \corresponds 8.3 \; \rm{\mu Jy/beam})$. The half-power beam width is $12\farcs67\times12\farcs08$ and the rms noise ($\sigma$) is $26~\rm{\mu Jy/beam}$. Contour levels are given by $\sigma\cdot(-3,3,6,12,24,48,96,192,384)$. Primary beam correction was applied.}}\label{6.2VLA}
\end{figure}

The galaxy NGC~4631 was observed for 9~h at 4.85~GHz in April 1999 with the VLA telescope in its D-array configuration (Project AD~896, here called data set VLA$_{1}$). These observations were pointed about $70\arcsec$ west of the dynamical center of NGC~4631. 
The data were calibrated and reduced in the AIPS data processing package. The nearby point source 1225+368 was used for phase and instrumental polarization calibration. The polarization position angle and flux-calibration were done by using 3C286 according to the flux scale published by \cite{Baars}. The UV data were self-calibrated for phase and amplitude. The maps of the three Stokes parameters I, U, and Q have a linear resolution of $12\farcs67\times12\farcs08$~HPBW. The polarized intensity and polarization angles were determined from the U- and Q-maps. The map of the total intensity together with the linear polarization is shown in Fig. \ref{6.2VLA}. The rms noise values are $26~\rm{\mu Jy/beam}$ in total intensity and $16~\rm{\mu Jy/beam}$ in polarized intensity.
In Table \ref{Tabledataavailabe} we list all VLA data available at these two frequency ranges, together with their distinct phase centers and beam sizes.

\begin{table}[h]
\begin{minipage}{1\columnwidth}
\begin{center}
\caption[Data used according to wavelength and telescope]{\small{Data used according to $\lambda$ and telescope, with its corresponding HPBW and pointing position (P.P.). C = near dynamical center of NGC~4631; NE = northeast of dynamical center; W = west of dynamical center; S.P.= Stokes Parameter.}}
\label{Tabledataavailabe}
\begin{adjustwidth}{}{-2.5em}
\begin{tabularx}{1\columnwidth}{|c|*8{>{\centering\let\\=\tabularnewline}X|}}
\hline
\multicolumn{1}{|c|}{\multirow{2}{*}{\textbf{$\bm \lambda$}}}& \multicolumn{1}{c|}{\multirow{2}{*}{\textbf{S.P.}}} & \multicolumn{2}{c|}{\textbf{Effelsberg}} & \multicolumn{2}{c|}{\textbf{VLA$_{\bm 1}$}} & \multicolumn{2}{c|}{\textbf{VLA$_{\bm 2}$}}\\
&  & \multicolumn{1}{c}{\textbf{\tiny{HPBW}}} & \multicolumn{1}{c|}{\textbf{P.P.}} & \multicolumn{1}{c}{\textbf{\tiny{HPBW}}}& \multicolumn{1}{c|}{\textbf{P.P.}} & \multicolumn{1}{c}{\textbf{\tiny{HPBW}}}& \multicolumn{1}{c|}{\textbf{P.P.}}\\ \hline \hline
\multirow{3}{*}{3.6 cm} & I & 85$''$ & C & - & - & 12$''$ & NE \\
                        & Q & 85$''$ & C & - & - & 12$''$ & NE \\
                        & U & 85$''$ & C & - & - & 12$''$ & NE \\ \hline
\multirow{3}{*}{6.2 cm} & I & 147$''$& C & \scriptsize{$12\farcs67\times12\farcs08$} & W & 12$''$ & NE \\
                        & Q & 147$''$& C & \scriptsize{$12\farcs67\times12\farcs08$} & W & - & -  \\

                        & U & 147$''$& C & \scriptsize{$12\farcs67\times12\farcs08$} & W & - & -  \\ \hline
\end{tabularx}
\end{adjustwidth}
\end{center}
\end{minipage}
\end{table}

\subsection{Merging of Effelsberg and VLA maps}

An additional VLA total intensity map of NGC~4631 at $\lambda$~6.2~cm was available from \cite{Golla}, hereafter called data set VLA$_{2}$. These observations were conducted in D-array configuration with 7.2~h of net observing time and were pointed northeast of the dynamical center of NGC~4631. 
Since \cite{Golla} used a zero-spacing flux density for the imaging of the total intensity map, the same was done for the imaging of our Stokes I VLA$_{1}$ map so that these two maps could be combined into a mosaic. 
With prior regridding and smoothing, the two VLA total intensity maps were mosaiced using the AIPS task LTESS. This task already applies the primary beam correction (PBC), which corrects for the primary beam attenuation with a cut-off parameter which we set to 25\% of the beam sensitivity at the phase center.

The Effelsberg and the VLA data at $\lambda$~6.2~cm were merged in the Fourier plane with the AIPS task IMERG. About 25\% of the total flux density of NGC~4631 given by the Effelsberg map was recovered with this procedure. The u,v-overlap adopted (UVRANGE) between the two data sets lies within 0.725~$\rm{k\lambda}$ and 0.775~$\rm{k\lambda}$. This range was determined based on two criteria. Firstly, the normalizing scale factor derived by AIPS for the combination should approximate the resolution difference between the two data sets. Secondly, the integrated flux density of the Stokes I merged map should be similar to that of the Effelsberg map.

The object at position $\alpha_{2000}=12^h41^m52^s; \delta_{2000}= 32^{\circ}27'00''$ is identified as a background source according to \cite{Hummel+Dettmar}. With our Effelsberg maps we determined that it has a flux density of 6.5~mJy (8~mJy) at 3.6~cm (6.2~cm). We tried to extract this bright source in both high- and low-resolution maps before merging them. However, in the low-resolution maps it is difficult to separate the flux of this source from the extended flux of NGC~4631 because of the size of the beams. Thus, this extraction heavily distorted our merged maps causing negative bowls to appear below the galaxy.

Data from the VLA at $\lambda$~3.6~cm (VLA$_{2}$) were available from \cite{Golla}. The phase center was chosen to be located northeast of the dynamical center of NGC~4631, as Golla and Hummel sought to study the radio spur in the northeastern halo. Prior to merging with the single-dish data, the interferometer maps were corrected for the primary beam attenuation and cut at 25\% of the sensitivity level at the beam center. For the merging of the data at this frequency the UVRANGE was between 1.1 and $1.2~\rm{k\lambda}$. In this case we recovered about 20\% of the flux density of NGC~4631 found in the Effelsberg map within the area of the VLA primary beam.

Consequently, we produced maps in polarized intensity (I$_{pol}$), polarization angle ($\psi$), and polarization degree (P) at both frequencies. The polarization angles were clipped at the $2\sigma$ level of the I$_{pol}$ map. The polarization degrees were clipped at the $2\sigma$ level of both the polarized intensity and the total intensity map. In addition, we smoothed our merged maps in all Stokes parameters to a resolution of 25$''$ and produced maps of the I$_{pol}$, $\psi$, and P at this angular resolution. The noise values of the maps are shown in Table \ref{tablenoise3.6}. Values for each map were estimated by calculating the average noise around the galaxy in small areas carefully selected to be free of emission. 

\begin{table}[h]
\begin{minipage}{1\columnwidth}
\centering 
\caption[Noise values in maps according to wavelength]{\small{Noise values according to $\lambda$ for maps at an angular resolution of $12\farcs67\times12\farcs08$~HPBW, with primary beam correction.}}
\begin{center}
\begin{tabular}[width=1\columnwidth]{ccc} 
\hline 
\textbf{Map} & \textbf{$\bm \lambda$~3.6~cm} & \textbf{$\bm \lambda$~6.2~cm} \\ 
\hline\hline 
I merged [$\rm{\mu Jy/beam}$]& 20 & 23\\
Q merged [$\rm{\mu Jy/beam}$]& 15 & 13   \\
U merged [$\rm{\mu Jy/beam}$]& 14 & 14 \\
I$_{pol}$ [$\rm{\mu Jy/beam}$] & 15 & 14 \\ 
\hline 
\end{tabular}
\end{center}
\label{tablenoise3.6} %
\end{minipage}
\end{table}

\section{Results}
\label{Results}

\subsection{Distribution of the total intensity emission}
\label{TPemission}

\begin{figure*}
\centering
\includegraphics[angle=90,width=1\columnwidth,angle=270]{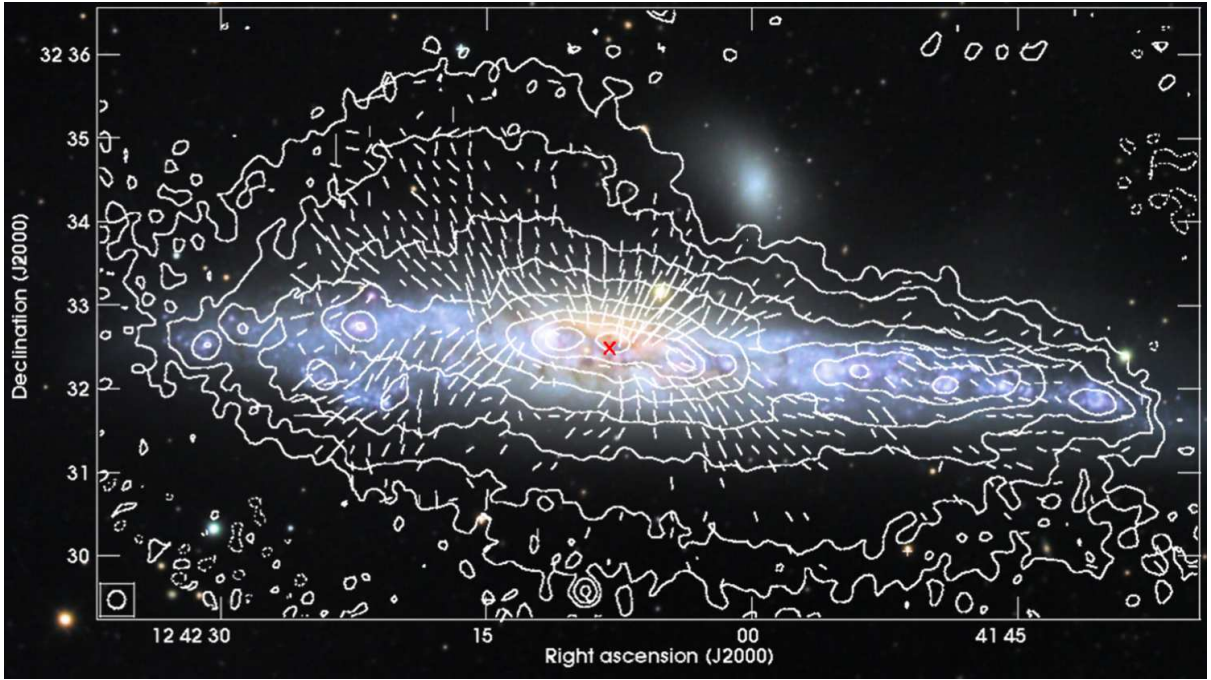}
\caption[]{\small{Total radio intensity map at 4.85~GHz (Eff+VLA) in contours with apparent magnetic field orientation. The length of the vectors is proportional to the I$_{pol}$ ($1''\corresponds 5.5 \; \rm{\mu Jy/beam} $). It is  overlayed on a colorscale optical DSS image (blue band). The half-power beam width is $12\farcs67\times12\farcs08$ and the rms noise ($\sigma$) is $23~\rm{\mu Jy/beam}$. Contour levels are given by $\sigma\cdot(-3,3,6,12,24,48,96,192,384)$. The ``x'' indicates the dynamical center of NGC~4631.}}
\label{6.2TP+E+90optical}
\end{figure*}

\begin{figure*}
\begin{center}
\includegraphics[angle=270,trim = 2mm 0mm 0mm 0mm, clip, width=1.8\columnwidth]{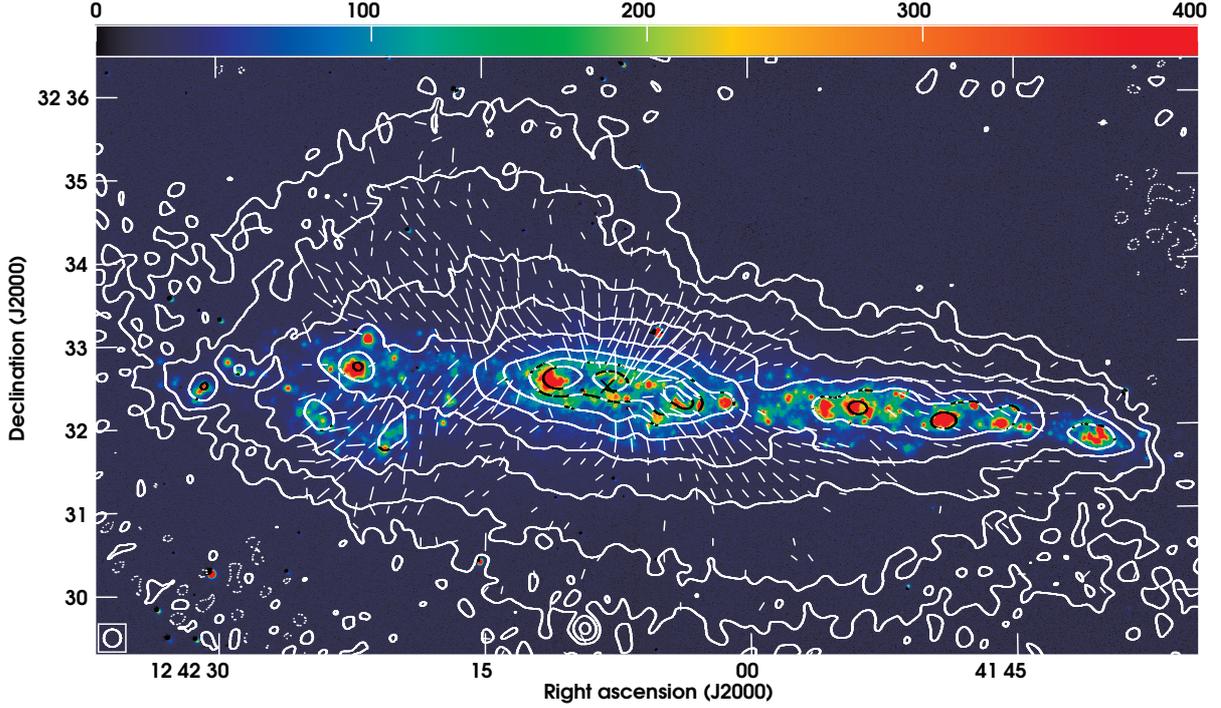}
\caption[]{\small{Total radio intensity map at 4.85~GHz (Eff+VLA) in contours with apparent magnetic field vectors. It is overlayed on a colorscale H$\alpha$ image. The length of the vectors is proportional to the I$_{pol}$ ($1'' \corresponds 5.5\; \rm{ \mu Jy/beam} $). The angular resolution is $12\farcs67\times12\farcs08$~HPBW and the rms noise ($\sigma$) is $23~\rm{\mu Jy/beam}$. Contour levels correspond to $\sigma\cdot(-3,3,6,12,24,48,96,192,384)$.}}\label{6.2TP+E+90Ha}
\end{center}
\end{figure*}

The maps of the VLA or merged total intensity emission of NGC~4631 at 4.85~GHz and 8.35~GHz are presented in contours in Figs. \ref{6.2VLA}, \ref{6.2TP+E+90optical}, \ref{6.2TP+E+90Ha}, \ref{3.6PIoverTP}, \ref{6.2PIoverTP}, and \ref{3.6B2}. At these radio frequencies the extent of NGC~4631 on the sky is larger than the primary beam of the VLA antennas. Therefore, the $\lambda$~3.6~cm total intensity map shown in contours in Fig. \ref{3.6PIoverTP} is limited by the primary beam size. Hence, we can only observe the eastern and central regions of NGC~4631 at this wavelength. On the other hand, thanks to the mosaic created with the high-resolution Stokes I maps at 4.85~GHz we were able to map the total intensity emission of NGC~4631 in its entire extent at this radio frequency (Figs. \ref{6.2TP+E+90optical}, \ref{6.2TP+E+90Ha}, and \ref{6.2PIoverTP}).

In order to estimate the total flux density of NGC~4631 we integrated the area of the galaxy in ellipses in the Effelsberg maps, obtaining a value of (430~$\pm$20)~mJy at $\lambda$~6.2~cm and (310~$\pm$~16)~mJy at $\lambda$~3.6~cm. 
The zero-level error and the error due to the noise in the integrated area contribute to the uncertainties in the determination of the flux density \citep{Noise}. In addition, we also considered the error due to the uncertainties in the integration parameters (radius and inclination of ellipses).

The radio emission of NGC~4631 along the major axis correlates well with the optical disk (Fig. \ref{6.2TP+E+90optical}) and H$\alpha$ emission (Fig. \ref{6.2TP+E+90Ha}). The dominant features in the total intensity maps are the extended central emission and the huge and asymmetric radio halo. The bright central emission extends $\sim$7~kpc along the major axis. The triple radio peaks (see Sect. \ref{intro}) are clearly visible at $\lambda$~6.2~cm. The IR center at 2~$\rm{\mu m}$, represented by the ``x'', lies within our central radio peak of the three; the westernmost coincides with the huge star forming region CM~67.

The $\lambda$~6.2~cm total intensity map has an asymmetric intensity distribution that is characteristic of this galaxy. The radio emission extends more to the west than to the east with respect to the dynamical center. Most prominent is the large z-extent of the halo in the northeast extending to about $3'$ above the major axis. The southern part of the halo is also asymmetric in shape with the largest extent in the southeast. Even the single dish Effelsberg map at $\lambda$~6.2~cm (Fig. \ref{6.2EFF}) shows a corresponding asymmetry in the northern part while the radio emission in the southern half is smoothed with the emission of the background source at this low angular resolution of $147''$~HPBW. The northern asymmetric distribution looks similar also at $\lambda$~20~cm by \cite{Hummel+Beck} and in the L-band regime by \cite{Westerbork}. However, a second larger extent of the radio emission is visible at $\lambda$~3.6~cm (Fig. \ref{3.6EFFTP}) in the northwestern halo which is already indicated in the $\lambda$~2.8~cm map by \cite{Dumke2.8}.

\subsection{Total and nonthermal spectral indices}
\label{sp}

With the integrated total flux-densities of NGC~4631 shown in Table \ref{tablespectra}, we fit a power law ($S_{tot} \; \propto \; \nu^{\alpha_{tot}}$) to calculate the integrated total spectral index of this galaxy. The flux density at 1.365~GHz was taken from \cite{SingsI} and at 10.55~GHz from \cite{Dumke2.8}. These values were incorporated because they are the most recent total flux-densities published of NGC~4631. The integrated total spectral index obtained is $\alpha_{tot}= -0.78 \pm 0.04$; it is in agreement with that published by \cite{Hummel+Dettmar}. This spectral index accounts for both the thermal and the nonthermal components of the radio emission.

However, the distinct spectral behavior of each of the two components can be used to distinguish one from the other. The thermal emission is considered to have a spectral index of $\alpha_{th}=-0.1$ ($S_{th} \; \propto \; \nu^{-0.1}$). The thermal fraction can be estimated with respect to that measured at another frequency as follows:
\begin{equation}
\label{frac}
{f_{1th}} = {{f_{2th}}\cdot{\left({\nu_1} \over {\nu_2}\right)} ^{-0.1-\alpha_{tot}}}.
\end{equation}

\begin{figure}[h!]
\centering
\includegraphics[trim = 2mm 0mm 0mm 0mm, clip, angle=270, width=1\columnwidth]{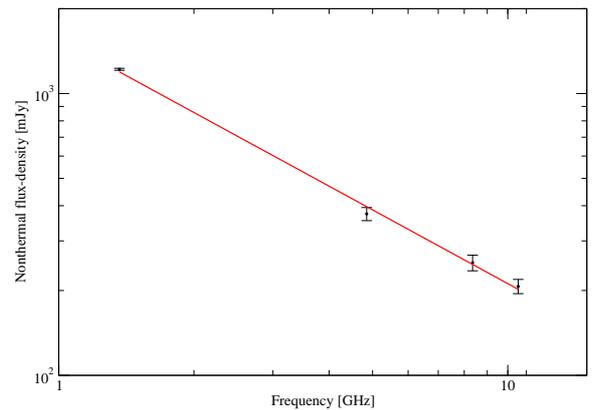}
\caption[Nonthermal integrated radio spectrum]{\small{Integrated radio spectrum of the nonthermal emission of NGC~4631. The red solid line shows the best-fit power law, with a nonthermal spectral index of $\alpha_{nth}= -0.87 \pm 0.03$.}}
\label{spec}
\end{figure}

By adding the approximate thermal contributions of the radio and $H\alpha$ emission, \cite{Gollaalone} estimated that NGC~4631 has a total thermal flux density at C-band (4.5-5.0~GHz) of 56~mJy. With this value we deduced a thermal fraction at 6.2~cm of 13\%. In Table \ref{tablespectra} we show our derived thermal fractions for each of the given frequencies. With the thermal fractions we derived the nonthermal flux-densities to estimate $\alpha_{nth}$. In Fig. \ref{spec} the solid red line represents the best fit to the power law resulting in a nonthermal spectral index of $\alpha_{nth}= -0.87 \pm 0.03$.

\begin{table}[h]
\begin{minipage}{1\columnwidth}
\caption[]{\small{Total flux-densities and thermal fractions ($f_{th}$) of NGC~4631 at four different frequencies.}}
\centering 
\begin{tabularx}{1\columnwidth}{c*4{>{\centering\let\\=\tabularnewline}X}}
\hline 
\textbf{Frequency [GHz]} & \textbf{Total flux density [mJy]} & $\rm{\bm f_{\bm t\bm h}}$  \\ 
\hline\hline 
1.365 & $1290\pm10$ \footnote{From \cite{SingsI}.} & 5.5\%  \\
4.85 & $430\pm20$ & 13\%\\
8.35 & $310\pm16$ & 19\%\\
10.55 & $265\pm12$ \footnote{From \cite{Dumke2.8}.} & 22\% \\
\hline
\end{tabularx}
\label{tablespectra} 
\end{minipage}
\end{table}

\subsection{Vertical scale heights of the total intensity emission}
\label{scaleheights}

While the observed vertical extent of the radio emission of an edge-on galaxy depends on the sensitivity of the radio map (i.e., the signal-to-noise ratio), the vertical scale height of the emission is a more appropriate parameter to describe and compare the vertical emission profile of edge-on galaxies. Of course, the fitting procedure has to take into account the telescope beam size and the inclination of the galaxy. This was done by introducing an {\em effective} beam size in the following way. We determined the intensity distribution in NGC~4631 along the major axis of the merged total intensity map at $\lambda$~6.2~cm and projected it to the inclination of the galaxy of $86\degr$. This distribution was convolved with a Gaussian function with $12\arcsec$ HPBW to simulate the effect of the telescope beam. The HPBW of the resulting distribution is the effective beam size. For the merged $\lambda$~6.2~cm map this is $16\arcsec$.

We determined the vertical scale heights at $\lambda$~6.2~cm from emission profiles that were obtained by a flux integration along three strips perpendicular to the major axis with a width of 150\arcsec{} each, centered on the nucleus. These profiles were fitted with model distributions (for details see \citealt{Dumke2.8}) consisting of an intrinsic one- or two-component exponential, or a Gaussian profile convolved with the effective beam size. The fits were made separately for the emission above (north = n) and below (south = s) the disk midplane (from east to west). 

\begin{figure*}[tbph]
\centering
\includegraphics[trim = 40mm 0mm 0mm 0mm, clip, angle=270, width=1.8\columnwidth]{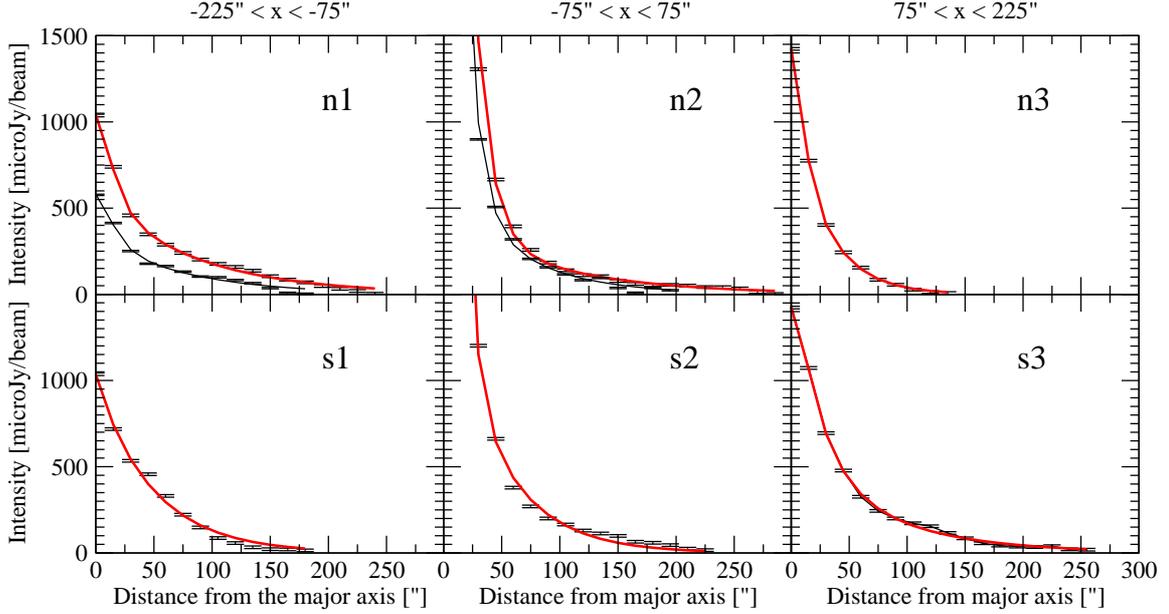}
\vspace{10pt}
\caption[]{\small{Total radio intensity profiles of the merged radio maps in $\rm{\mu Jy/beam}$ perpendicular to the major axis of NGC~4631. The measured points are averaged in strips of 150\arcsec width along the major axis. The x-axe of the plots give the distance from the major axis in [\arcsec]. The upper panels refer to the northern part (i.e., above the major axis) and the lower panels to the southern part (i.e., below the major axis) of NGC~4631. The thick red lines represent the two-component exponential fit to the data at 4.85~GHz; the thin black lines in n1 and n2 represent the fit to the 8.35~GHz data.}}
\label{xmplot}
\end{figure*}

As the merged $\lambda$~3.6~cm map only covers the northeastern part of NGC~4631 fully, we could only determine and fit the flux distributions for n1 and n2 at this wavelength. A two-component exponential function fitted the data best at both wavelengths as was already found for several other spiral galaxies seen edge-on \citep{Krause2011}.  The emission profiles and their errors together with the two-component fits to the profiles are shown in Fig. \ref{xmplot}. The corresponding values for the scale heights for the thin and thick disks are summarized in Table \ref{scale}.

\begin{table}[h]
\caption{\small{Vertical scale heights for the thin and thick disk. n = north and s = south. The eastern (1), central (2) and western (3) strips have a width of 150\arcsec each.}}
\label{scale}
\begin{center}
\begin{tabular}[]{crrcrcrc}
\hline
& \multicolumn{1}{c}{\textbf{Range along}} &\multicolumn{2}{c}{\textbf{8.46~GHz}}&\multicolumn{2}{c}{\textbf{4.86~GHz}}\\
\textbf{Strip} & \textbf{major axis} & $\rm{\bm h_{\bm t\bm h\bm i\bm n}}$ & $\rm{\bm h_{\bm t\bm h\bm i\bm c\bm k}}$ & $\rm{\bm h_{\bm t\bm h\bm i\bm n}}$ & $\rm{\bm h_{\bm t\bm h\bm i\bm c\bm k}}$\\
& \multicolumn{1}{c}{\textbf{[$\bm \arcsec$]}} & \textbf{[pc]}&\textbf{[kpc]}&\textbf{[pc]}&\textbf{[kpc]}\\
\hline \hline
n1 & -225 to~~-75 & 420 & 2.9 & 440 & 3.2\\
n2 &  -75 to~~~75 & 410 & 2.2 & 490 & 3.4\\
n3 &   75 to~225 &  & & 160 & 1.1\\
s1 & -225 to~~-75 & &  &  30 & 1.8\\
s2 &  -75 to~~~75 & & & 290 & 1.7\\
s3 &   75 to~225 & & & 810 & 2.9\\
\hline
\textbf{Mean} & & & & $370\pm 280$ & $2.3\pm 0.9$\\
\hline
\end{tabular}
\end{center}
\end{table}

The fitted values for the scale heights vary widely within the different strips in NGC~4631 at $\lambda$~6.2~cm for both the thin and thick disk, unlike the values for other galaxies. This is also reflected by the large standard deviations of the mean values (see Table \ref{scale}). The mean values themselves are higher than the values found for all of the other six edge-on spiral galaxies observed so far: $300\pm50$~pc for the thin disk and $1.8\pm0.2$~kpc for the thick disk (\citealt{DumMK98}, \citealt{Krause2011}).

Within NGC~4631 the scale heights in n1, n2, and s3 have significantly larger values than in other spiral galaxies. Those in n3 and the thin disk in s1 have smaller values. Only the thick disk in s1 and both quantities in s2 have similar values to those found in other spiral galaxies. The emission does not show the typical dumbbell structure visible in NGC~253, for example \citep{NGC253}, and in NGC~4565 \citep{Krause2009}. The dumbbell structure with its smaller extent of the emission above and below the central region of the disk and larger z-extents at larger radii seems to reflect dominating synchrotron losses in a magnetic field that is strongest along the \emph{central} galactic plane of the galaxy \citep{NGC253}. The galaxy NGC~4631 is different in this respect. This will be further discussed in Sect.~\ref{Discussion}.

\subsection{Distribution of the polarized emission and degree of polarization}
\label{PI}

\begin{figure}[h]
\centering
\includegraphics[trim = 2mm 0mm 0mm 0mm, clip, width=0.6\columnwidth, angle=270]{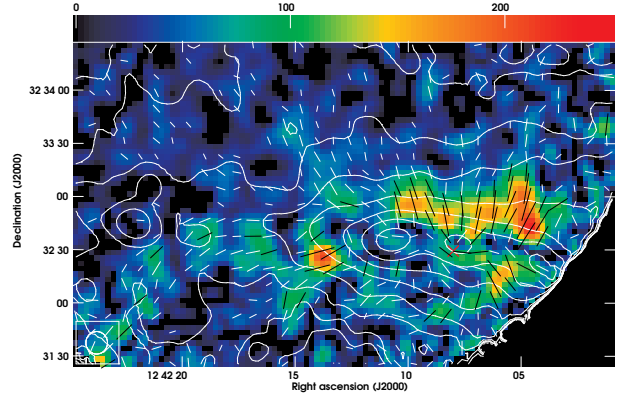}
\caption[]{\small{Colorscale of the polarized emission over total radio intensity contours at $\lambda$~3.6~cm (Eff+VLA), with apparent magnetic field vectors. The length of the vectors is proportional to the I$_{pol}$ ($1'' \corresponds 9.5\; \rm{\mu Jy/beam}$). The colorscale gives the polarized intensities in $\rm{\mu Jy/beam}$. The half-power beam width is $12\farcs67\times12\farcs08$. Contour levels of the total radio emission correspond to $(20~\rm{\mu Jy/beam})\cdot(-3,3,6,12,24,48,96,192,384)$.}}
\label{3.6PIoverTP}
\end{figure}

\begin{figure}[h]
\centering
\includegraphics[trim = 2mm 0mm 0mm 0mm, clip,width=0.55\columnwidth, angle=270]{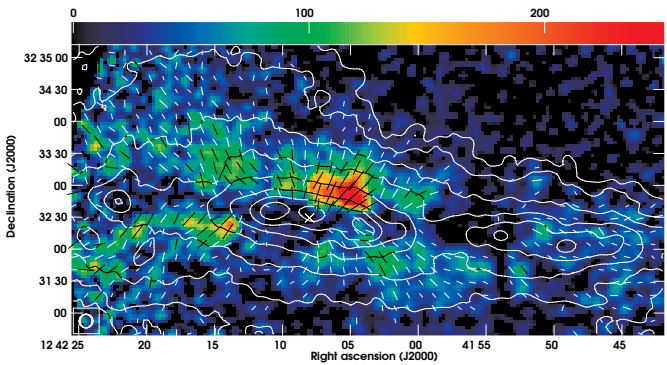}
\caption[]{\small{Colorscale of the polarized emission over total radio intensity contours at $\lambda$~6.2~cm (Eff+VLA), with apparent magnetic field orientation. The length of the vectors is proportional to the I$_{pol}$ ($1'' \corresponds 8.3\; \rm{\mu Jy/beam }$). The colorscale gives the polarized intensities in $\rm{\mu Jy/beam}$. The half-power beam width is $12\farcs67\times12\farcs08$. Contour levels of the total radio emission are given by $(23~\rm{\mu Jy/beam})\cdot(-3,3,6,12,24,48,96,192,384)$.}}
\label{6.2PIoverTP}
\end{figure}

The linearly polarized emission of NGC~4631 at 8.35~GHz and 4.85~GHz is plotted as a colorscale in Figs. \ref{3.6PIoverTP} and \ref{6.2PIoverTP} with a resolution of $12\farcs67\times12\farcs08$~HPBW. 

There is evidently more polarized emission above than below the plane of NGC~4631. At $\lambda$~6.2~cm the emission extends $\sim1\farcm6$ north of the plane and $\sim1\farcm2$ south of the plane. In addition, the X-ray emission \citep{Wang} is also brighter on the northern part of the galaxy. This implies that we might be observing NGC~4631 from above its disk. As expected, at $\lambda$~3.6~cm the polarized emission is more concentrated towards the major axis because of the decrease of Faraday depolarization effects at shorter wavelengths.

The polarized emission of NGC~4631 can be separated into four extra-planar quadrants with respect to the eastern peak of the triple radio source (see Sect. \ref{TPemission}) at position $\alpha_{2000}=12^h42^m11^s; \delta_{2000}= 32^{\circ}32'30''$. The northwestern quadrant has the brightest polarized intensity at both wavelengths, followed by the southeastern quadrant. At $\lambda$~3.6~cm in Fig. \ref{3.6PIoverTP}, also visible somewhat broader at $\lambda$~6.2~cm in Fig. \ref{6.2PIoverTP}, there is an inner spur-like feature distinguishable towards the northeast of the galactic center. This structure was already studied by \cite{Golla}, who state that it is a small tributary of the northeastern spur visible in total power (see Fig. \ref{3.6EFFTP}) and that it seems to originate from the eastern peak of the triple source (CM67).

\begin{figure}[h]
\centering
\includegraphics[trim = 2mm 0mm 0mm 0mm, clip, angle=270,width=0.95\columnwidth]{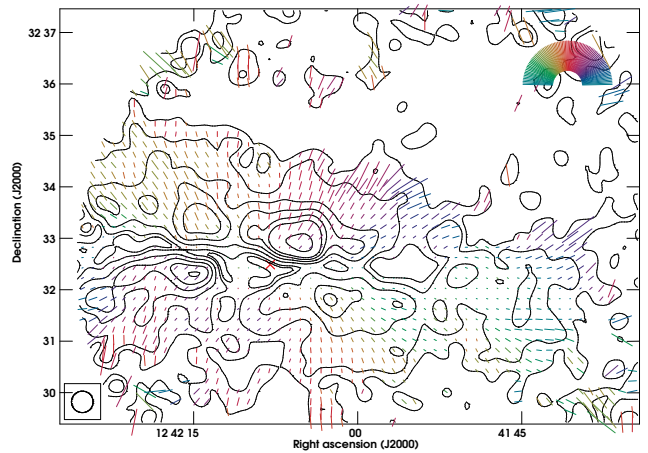}
\caption[]{\small{Linearly polarized emission at $\lambda$~6.2~cm (Eff+VLA), with apparent magnetic field orientation. The length of the vectors is proportional to the polarization degree ($1'' \corresponds 2\% $). The angular resolution is $25''$ and the rms noise ($\sigma$) is 20~$\rm{\mu Jy/beam}$. Contour levels are given by $\sigma\cdot(-2,2,4,8,12,16,20,32)$. The ``x'' indicates the dynamical center. In the electronic version the angles are color coded as shown in the colorfan in the upper-right corner.}}\label{6.2PI}
\end{figure}

In order to analyze the large-scale structure and increase the signal-to-noise ratio, the Stokes Q- and U-maps were smoothed to a resolution of $25''$ and polarized intensities were determined from these (see Fig. \ref{6.2PI}). 
Again, there are four peaks of polarized emission distributed around the dynamical center of NGC~4631, in each of the four quadrants previously mentioned. 

The polarized intensity shows a minimum along the galactic midplane at both wavelengths. This is expected as Faraday depolarization effects are usually strongest in this area because the thermal electron density, the LOS, and the magnetic field strength are largest. A depolarization along the midplane has already been observed in several other spiral edge-on galaxies, for example NGC~5775 \citep{Soida2011}. Furthermore, if there is a magnetic field parallel to the disk accompanied by strong vertical magnetic fields above and below the disk, we also expect strong beam polarization along the region of the projected transition between both magnetic field components. To minimize the beam depolarization, observations with high resolution are required, while high frequency observations reduce the Faraday depolarization. We even found that the merging of single-dish data to the interferometer data increases the zone of depolarization along the midplane of NGC~4631 (see Fig. \ref{6.2PIoverTP}). This is expected along the transition zone of two large-scale magnetic field patterns with different orientations: the beam depolarization in the single-dish maps is much larger and reduces the observed U- and Q-signal at positions where the interferometer beam still detects a signal. In these cases the merged map shows less polarized intensity than the original interferometer map (see Fig. \ref{6.2PIoverTP}).

\subsection{Rotation measure distribution and the magnetic field along the LOS}
\label{RM}

The polarization angles can only be determined with an $n\cdotp\pi$ ambiguity which leads to an uncertainty in the $RM$ derivation. For $n=1$, at wavelengths of 3.6 and 6.2~cm, we would have ambiguous rotation measure values of: $RM=\pi/({\lambda_{2}}^2-{\lambda_{1}}^2)\approx\pm1230~\rm{rad/m^2}$, which is high compared to the expected values for a galactic disk.

The obtained rotation measures are biased by the Galactic foreground component ($RM_{fg}$), which is quite small in the direction of the sky in which NGC~4631 is located. Nevertheless, we corrected for this by subtracting the value of $RM_{fg}=(-4\pm3)~\rm{rad/m^2}$ \citep{Westerbork} from our rotation measures.

The overall $RM$ distribution obtained does not present values that exceed $\pm615~\rm{rad/m^2}$. Even without being affected by the $n\cdotp\pi$ ambiguity, a rotation of the polarization angle by 90$^\circ$ along the LOS through the emitting source leads to strong differential Faraday depolarization within the galaxy. This is expected at $\lambda$~6.2~cm for rotation measure values of $|RM|\approx|(\pm\pi/2)/({6.2\cdot10^{-2}})^{2}|\approx400~\rm{rad/m^2}$ and means that the galaxy is no longer transparent in polarized emission. Hence, this implies that the derived values for the rotation measure may be incorrect. To be on the safe side we set our rotation measure threshold to $|RM|~\leq~350~\rm{rad/m^2}$ and clipped both the rotation measure distribution and the polarization angles to satisfy this condition. We determined the rotation measure distribution at $25''$ with the original VLA maps at $\lambda$~6.2~cm and 3.6~cm. This map is used to correct the polarization angles observed in Fig. \ref{3.6B2}.

\begin{figure}[h]
\centering
\includegraphics[trim = 2mm 2mm 2mm 2mm, clip,width=0.85\columnwidth,angle=270]{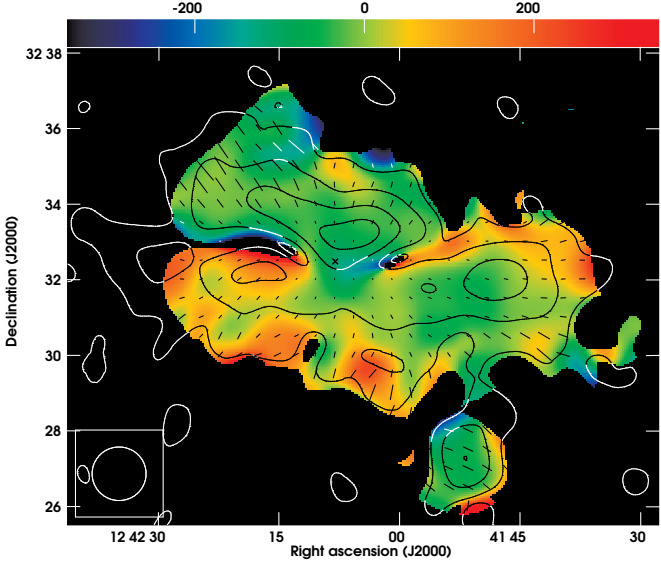}
\caption[]{\small{Rotation measure distribution colorscale (6.2cm-Merged+3.6cm-Eff) over Effelsberg $\lambda$~3.6~cm I$_{pol}$ contours, with intrinsic magnetic field. Length of the vectors is proportional to the Effelsberg $\lambda$~3.6~cm polarization degree ($1'' \corresponds 1\%$). These vectors were clipped at the $2\sigma$ level of the polarized intensity maps (at both frequencies) and where $|RM|~\geq~350~\rm{rad/m^2}$. The colorscale gives the rotation measures in $\rm{rad/m^2}$. All values are from the data at an angular resolution of $85''$~HPBW. Contour levels of the Effelsberg $\lambda$~3.6~cm polarized emission are given by $(62~\rm{\mu Jy/beam})\cdot(-3,3,8,16,24)$. }}\label{2RM25}
\end{figure}

To analyze the large-scale rotation measure pattern of NGC~4631 we determined the rotation measures between the  Effelsberg $\lambda$~3.6~cm map (Fig. \ref{3.6EFFTP}) and the merged $\lambda$~6.2~cm map smoothed to $85''$ resolution. In Fig. \ref{2RM25} we show these rotation measures with respect to the Effelsberg polarized emission at $\lambda$~3.6~cm ($85''$~HPBW). Overall, the rotation measure distribution is smooth and regular. The concentration of the thermal emission towards the disk, due to the elevated electron densities in the high starforming regions, may explain why $|RM|$ increases close to the plane and especially around the dynamical center. We note that the rotation measure in the northeast quadrant is negative with increasing $|RM|$ values from the midplane in a region around the northeastern spur (see Sect. \ref{PI}). This region, which we call the \emph{Northeastern extension}, will be discussed further in Sect. \ref{Discussion}.

\subsection{Intrinsic magnetic field orientation perpendicular to the LOS}
\hspace{1 cm}
\label{B-FIELD}
The apparent magnetic field orientation (which are the observed polarization angles rotated by $90^{\circ}$) are shown as vectors in the merged maps in Figs. \ref{6.2TP+E+90optical}, \ref{6.2TP+E+90Ha}, \ref{6.2PIoverTP}, and \ref{6.2PI} at $\lambda$~6.2~cm and in Fig. \ref{3.6PIoverTP} at $\lambda$~3.6~cm. In order to obtain the intrinsic magnetic field orientation in NGC~4631, the observed polarization angles at $\lambda$~3.6 and 6.2~cm were corrected for Faraday rotation with the rotation measure distribution. Rotating the intrinsic polarization angles by $90^{\circ}$ yields the orientation of the magnetic field component perpendicular to the LOS (B$_\perp$-vectors). It is worth noting that the polarized intensity is only sensitive to the field orientation but not to the field direction, hence it cannot distinguish between parallel and antiparallel field directions in the plane of the sky.

We determined the vectors at those locations where the polarized intensity at both wavelengths is above $2\sigma$ and where the rotation measures are below $|RM|~\leqslant~350~\rm{rad/m^2}$, which is where the derived values for the rotation measures are reliable, as discussed in Sect. \ref{RM}. By error propagation the error in the intrinsic polarization angles is ${\sigma_{\psi int}}^2= {\sigma_{\psi}}^2 + (\lambda^2 \cdot \sigma_{RM})^2 $. Hence, in areas where the polarized emission is twice the rms noise ($\sigma_{I_{pol}}\approx \sigma_{Q,U}$), the error in the intrinsic polarization angles is at 6.2~cm $\sim34^\circ$ and at 3.6~cm $\sim18^\circ$. However, this error scales down linearly with increasing polarized intensity. 

\begin{figure}[h]
\begin{center}
\includegraphics[width=0.55\columnwidth,angle=270, trim = 2mm 5mm 0mm 0mm, clip]{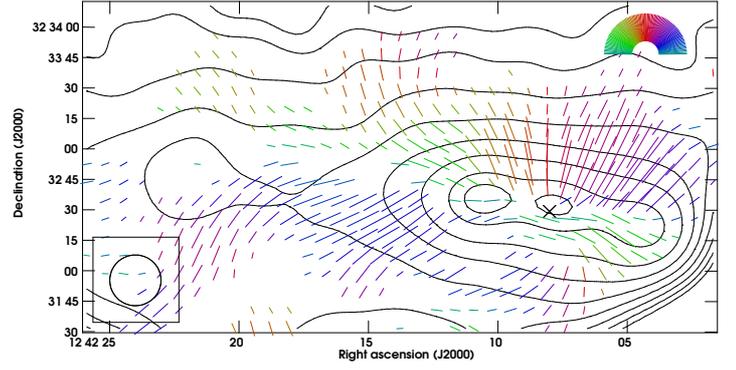}
\caption[]{\small{Intrinsic magnetic field orientation perpendicular to the LOS (B$_\perp$-vectors) derived from the 25$''$ resolution data (VLA only). The contours show the VLA $\lambda$~3.6~cm total radio emission. The length of the vectors is proportional to the VLA $\lambda$~3.6~cm polarized intensity ($1'' \corresponds 15\; \rm{\mu Jy/beam} $). These vectors were clipped at the $2\sigma$ level of the polarized intensity maps (at both frequencies) and where $|RM|~\geq~350~\rm{rad/m^2}$. The contour levels correspond to $(40~\rm{\mu Jy/beam})\cdot(-3,3,6,12,24,48,96,192,384,522)$. In the electronic version the angles are color coded as shown in the colorfan in the upper-right corner.}}\label{3.6B2}
\end{center}
\end{figure}

Merging with Effelsberg data is essential for recovering the large-scale emission of the total intensity and the linear polarization in NGC~4631. In special cases, however, the polarized signal may even be weakened by the merging process, e.g., along a transition zone of large-scale magnetic field patterns with abruptly changing field orientations (as discussed in Sect. \ref{PI}). At $\lambda$~6.2~cm we found that the merging with the single-dish data actually reduced the amount of polarization vectors along the disk (compare Figs. \ref{6.2VLA} and \ref{6.2PIoverTP}). Therefore, in order to minimize the beam depolarization effects along the midplane we determined the intrinsic magnetic field orientation derived solely with the VLA data at both frequencies. To increase the signal-to-noise ratio we smoothed these data to $25''$ and corrected them for Faraday rotation with the corresponding RM map (see Sect. \ref{RM}). The map of the intrinsic magnetic field in presented in Fig. \ref{3.6B2}.

We can clearly see that the halo magnetic field orientation in the eastern side of the galaxy is almost symmetric with respect to the major axis of NGC~4631 and that it curves radially outwards. At the edges of the western side of the maps we can also distinguish that the halo magnetic field begins to curve outwards. This forms part of the X-shaped halo magnetic field seen in other edge-on galaxies. As claimed by \cite{Golla}, along the spur-like feature observed in polarized emission (see Sect. \ref{PI}) the magnetic field orientation seems to follow this spur. In addition, above the dynamical center the halo magnetic field is orientated perpendicular to the major axis of the galaxy.

\begin{figure}[h]
\centering
\includegraphics[angle=270,trim = 2mm 2mm 2mm 2mm, clip, width=1\columnwidth]{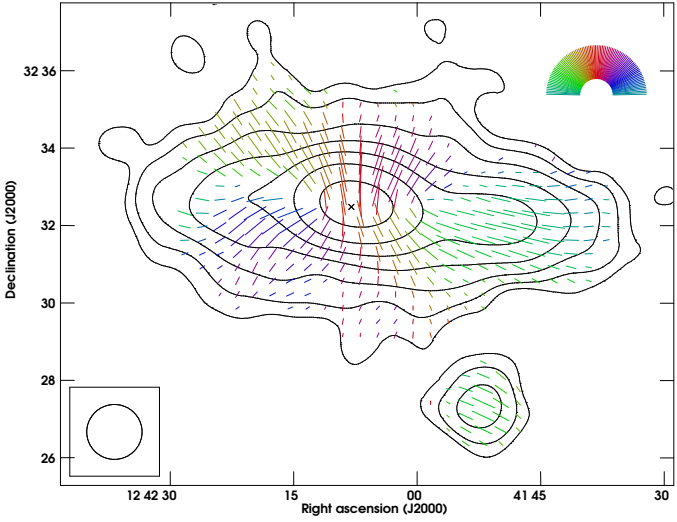}
\caption[]{\small{Intrinsic magnetic field orientation derived from the 85$''$ resolution data (6.2cm-Merged+3.6cm-Eff). Contours correspond to the Effelsberg $\lambda$~3.6~cm total radio emission. The length of the vectors is proportional to the Effelsberg $\lambda$~3.6~cm polarized intensity ($1'' \corresponds 26\; \rm{\mu Jy/beam} $). These vectors were clipped at the $2\sigma$ level of the polarized intensity maps (at both frequencies) and where $|RM|~\geq~350~\rm{rad/m^2}$. The contour levels correspond to $(350~\rm{\mu Jy/beam})\cdot(-3,3,6,12,24,48,96,192,384)$. In the electronic version the angles are color coded as shown in the colorfan in the upper-right corner.}}\label{85B}
\end{figure}

Along the central region of the disk the B$_\perp$-vectors are oriented parallel to the galaxy's plane extending over several beam sizes. We corrected the $\lambda$~3.6~cm Effelsberg map (Fig. \ref{3.6EFFTP}) for Faraday rotation with the RM map determined between this map and the merged $\lambda$~6.2~cm map smoothed to $85''$ (Fig. \ref{2RM25}). The map is shown in Fig. \ref{85B} and presents a plane-parallel magnetic field in the outer regions along the midplane that had previously been indicated by \cite{Krause2009}. With the plane-parallel vectors in Fig. \ref{3.6B2} we can now see for the first time the missing link of a disk-parallel magnetic field in the central 5-7~kpc in NGC~4631.

\subsection{Depolarization}
\label{depolarization}

\begin{figure}[h]
\centering
\includegraphics[trim = 2mm 2mm 2mm 2mm, clip,width=0.85\columnwidth,angle=270]{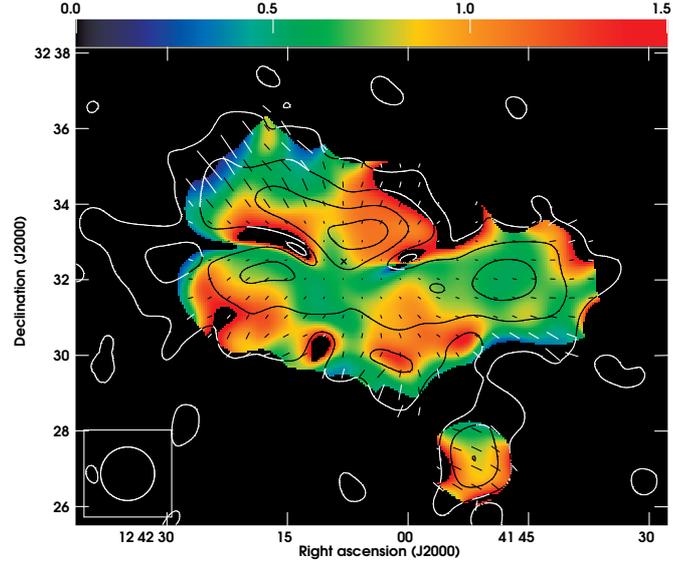}
\caption[]{\small{Colorscale of the depolarization distribution (6.2cm-Merged+3.6cm-Eff) over Effelsberg $\lambda$~3.6~cm I$_{pol}$ contours, with intrinsic magnetic field. Length of the vectors is proportional to the Effelsberg $\lambda$~3.6~cm polarization degree ($1'' \corresponds 1\%$). These vectors were clipped at the $2\sigma$ level of the polarized intensity maps (at both frequencies) and where $|RM|~\geq~350~\rm{rad/m^2}$. The contour levels correspond to $(350~\rm{\mu Jy/beam})\cdot(-3,3,6,12,24,48,96,192,384)$. All values are from the data at an angular resolution of $85''$~HPBW. Contour levels of the Effelsberg $\lambda$~3.6~cm polarized emission are given by $(62~\rm{\mu Jy/beam})\cdot(-3,3,8,16,24)$. }}\label{DEPOL}
\end{figure}

The observed Faraday depolarization (DP) (wavelength-dependent) in NGC~4631 was determined by the ratio of the polarization degree at $\lambda$~6.2~cm (merged data set) to that at 3.6~cm (Effelsberg data set) with a resolution of $85''$. Since the maps of the polarization degree at both frequencies have the same angular resolution, the beam depolarization is equal in both maps and cancels out, except for the effects along the midplane as discussed in Sect. \ref{PI}. The observed DP is presented in Fig. \ref{DEPOL}. It was determined from data truncated at $2\sigma$ of the I$_{pol}$ maps and at $3\sigma$ of the total intensity maps. The depolarization values in some regions are DP $> 1$; this may be due to additional errors in the maps caused by the large separation between the pointing positions of the different observations. 

As expected, the depolarization is higher in the plane of the galaxy which could be the result of several factors. The H$\alpha$ distribution of NGC~4631 (Fig. \ref{6.2TP+E+90Ha}) indicates that the thermal electrons are concentrated in the disk of the galaxy in the star forming regions. The galaxy NGC~4631 is known to have strong star formation activity (see Sect. \ref{intro}). Since star formation causes the magnetic field to become more turbulent, the interstellar medium in the disk may be considerably more turbulent producing internal Faraday dispersion. Furthermore, the long LOS through the disk may also cause depolarization by differential Faraday rotation within NGC~4631. In addition, rotation measure gradients due to the decrease of thermal electrons in z-direction can also produce depolarization, as estimated by \cite{Golla} for our frequency range.

The depolarization seems to increase towards the extent of the northeastern halo of NGC~4631. In the uppermost region of the northeastern halo, we also have high rotation measures (about $-150\;\rm{rad/m^2}$) as seen in Fig. \ref{2RM25}. This is surprising since in general the depolarization effects decrease towards the halo because of the decreasing thermal electrons and the decreasing LOS through the emission. Another possibility is that the B$_\shortparallel$-field in this region is stronger. This would be the case if the magnetic field lines are highly ordered and bend away from us along the spur. This would cause an increase in the B$_\shortparallel$-field and hence strong depolarization and high rotation measures.

\subsection{Magnetic field strengths}
\label{strength}

If one assumes equipartition between the total energy densities of cosmic rays and that of the magnetic field, the magnetic field strength can be derived from the nonthermal radio emission as described in \cite{Equipartition}. In Sect. \ref{B-FIELD}  we described the configuration of the B$_\perp$-field orientation of NGC~4631 to be composed of an X-shaped field in the halo, an extra-planar vertical field above the dynamical center and a field along the disk. We estimated the magnetic field strengths with the total and polarized emission of the Effelsberg $\lambda$~3.6~cm map in different regions within the halo and the disk of more than a beamsize each, according to the different field geometries in the disk and halo, assuming a constant nonthermal spectral index of $\alpha_{nth}= -0.87$ (see Sect. \ref{sp}). The estimated strengths for the total and ordered field are summarized in Table \ref{tablestrengths} for the different regions . The widths of the two regions along the disk are about 2~kpc each. The region with the vertical field above the dynamical center extends from about $1-4~$kpc above the central midplane and has a radial width of about 4~kpc. The regions in the four quadrants (N-E, S-E, N-W, S-W) are located symmetrically relative to the center at $\rm{|z|}\simeq 1$~kpc and  $\rm{|r|}\simeq 2$~kpc with a size of about 3.5~beams. For the region called the \emph{Northeastern extension} only the upper part of the northeastern halo was considered (between about $4-9$~kpc above the midplane); it also has a size of about 3.5~beams. The errors given in Table \ref{tablestrengths} include the measurement errors but are dominated by the uncertainties in $\alpha_{nth}$, the assumed LOS, and inclination of the field.

\begin{table}[h]
\begin{minipage}{1\columnwidth}
\caption[Magnetic Field Strengths]{\small{Magnetic field strengths in different areas of NGC~4631. $B_t$ - Total field strength. $B_{ord}$- Ordered field strength. }} 
\centering 
\begin{tabularx}{1\columnwidth}{c*4{>{\centering\let\\=\tabularnewline}X}}
\hline 
\textbf{Region} & \textbf{Area} & \textbf{B$_{\rm{\bm t}}~[\rm{\mu G}]$} & \textbf{B$_{\rm{\bm o\bm r\bm d}}~[\rm{\mu G}]$}  \\ 
\hline\hline 
Halo & N-E;S-E;N-W;S-W & $10\pm 2$ & $3\pm1$ \\
Halo & \emph{\small{Northeastern extension}} \footnote{see Sect. \ref{RM}.} & $7\pm2$ & $5\pm2$ \\
Halo & \small{Vertical field above dyn. center} & $11\pm2$ & $4\pm1$ \\ \hline
Disk & \small{Central 5~kpc along disk} & $13\pm2$ & $2\pm1$ \\ 
Disk & \small{Entire disk} & $9\pm2$ & $2\pm1$ \\ \hline
\end{tabularx}
\label{tablestrengths} 
\end{minipage}
\end{table}

Within the range of uncertainties, the total and ordered field strengths seem to be very similar in all of the considered regions of the X-shaped field. Furthermore, the strength of the total field in the central region of the disk is comparable to the strength of the perpendicular field above it within the margin of uncertainties. In addition, the strength of the total magnetic field in the entire disk seems to be similar to the strengths of the total field in the halo. The strengths of the ordered field in the halo seems to be at least as strong as that of the ordered field in the disk. The degree of uniformity (the ratio of ordered to total field strength) in the halo is generally higher than in the disk, especially along the \emph{Northeastern extension}. 

\section{Discussion}
\label{Discussion}

In Sect. \ref{B-FIELD} we described our results for the magnetic field orientation of NGC~4631 on the sky plane (B$_\perp$). In the halo it is characterized by an overall X-shaped configuration and a strong perpendicular field above the galactic center. The edge-on spiral galaxies NGC~5775 \citep{Soida2011} and NGC~4666 \citep{Soida2005} also have strong perpendicular components above and below their galactic centers. Dynamo simulations by \cite{MHD} may be able to explain not only this X-shaped morphology but also these vertical fields.

The Faraday-corrected single-dish Effelsberg map at $\lambda$~3.6~cm in Fig. \ref{85B} shows 
a plane-parallel magnetic field in the outer eastern and western regions along the midplane which was already indicated by \cite{Krause2009}. With the plane-parallel vectors in Fig. \ref{3.6B2} (corrected for Faraday rotation) we can now see for the first time the missing link of a disk-parallel magnetic field in the central 5-7~kpc in NGC~4631. Hence, NGC~4631 is no longer an outstanding object since it also contains a disk-parallel magnetic field along its midplane as has been detected in all of the other face-on and edge-on galaxies observed so far (e.g., \citealt{Beck+Wielebinski2013}). This plane-parallel magnetic field is thought to be generated by a large-scale $\alpha\Omega$~dynamo. The galaxy NGC~4631 was regarded as an exception, but our research indicates this is not the case.

The total magnetic field strength in the disk of NGC~4631, $B_{t}\approx9\pm2~\rm{\mu G}$, is comparable to the total field strength of the star forming edge-on galaxy NGC~5775 \citep{Soida2011}, but it is not as strong as the strength in the starburst galaxy NGC~253 \citep{NGC253} which also has strong star formation along its disk. However, the average strength of the total magnetic field in the halo, $B_t\approx10\pm2~\rm{\mu G}$, is higher than in other the edge-on galaxies studied so far. In addition, the strengths of the ordered field in the halo $B_{ord}\approx(3-5)\;\rm{\mu G}$ seem to be larger than the strength of the ordered field in the disk $B_{ord}\approx2\pm1\;\rm{\mu G}$ (see Table \ref{tablestrengths}). Hence, strong differential Faraday rotation in the disk as well as in the halo is expected. This, together with strong beam depolarization along the transition zone between the horizontal disk field and the mainly vertical halo field in the central region of NGC~4631, might explain why the plane-parallel field in this disk was so difficult to detect when compared to other galaxies.

Our rotation measure distribution between $\lambda$~3.6 and 6.2~cm of NGC~4631 (see Sect. \ref{RM}) shows a smooth variation over scales larger than the beam size. This indicates that there is indeed a large-scale, regular magnetic field (B$_\shortparallel$) in the halo of NGC~4631. As the RM distribution presents no type of symmetric pattern with respect to the major or minor axis, we cannot associate it with a specific parity of the presumably dynamo generated magnetic field in the disk. Close to the galactic plane the rotation measures are high ($|RM|\in[75,\ge 350]\;\rm{rad/m^2}$), as one would expect because of the long path-length through the entire galactic disk and the concentration of thermal emission.

In the northeastern halo of NGC~4631 the rotation measures increase farther upwards, reaching values of up to $-150\;\rm{rad/m^2}$. Since the thermal electron density decreases with increasing z-values \citep{Hummel+Beck}, this indicates that the strength of the B$_\shortparallel$ in the uppermost region is considerable. Assuming a bending of the field by about $40^\circ$ in this \emph{Northeastern extension} we calculated the total and ordered field strength and obtained values of $B_t\approx7\pm2~\;\rm{\mu G}$ and $B_{ord}\approx5\pm2~\;\rm{\mu G}$, while the total field strength without a bending is only $B_t\approx5~\;\rm{\mu G}$. The ratio of the ordered to the total field strength is highest here, i.e., the degree of uniformity of the magnetic field structure along the  \emph{Northeastern extension} is highest compared to other regions in the halo (and the disk). In addition, the increase of depolarization and the decrease of the polarized emission can also be explained by a bending of the magnetic field lines away from us.

The orientation of the halo B$_\perp$-field in the northeastern quadrant of NGC~4631 seems to be aligned with the HI-spur located in this region (see Sect. \ref{intro}). Since this is also about parallel to the X-shaped halo field, it is difficult to distinguish which part of the B$_\perp$-vectors might be related to the HI-spur and hence to determine if these features are actually related. This HI spur (labeled 4 in \citealt{Rand1994}) is part of an extended HI structure around NGC~4631 which is related to tidal interaction mainly between the three galaxies NGC~4631, NGC~4656, and NGC~4627. It has been argued already by \cite{Hummel} that this tidal interaction of NGC~4631 is responsible for the large extent of its radio halo (e.g., \citealt{Hummel+Dettmar}). According to \cite{Hummel} the magnetic field in the disk may have been pulled out of the disk together with the gas leading to strong ordered magnetic fields in the halo of NGC~4631. The large ordered magnetic field strength in the halo and their strong vertical components could now be verified by our observations at higher frequency. The smooth RM-pattern even indicates a coherent magnetic field structure in the halo.

Other huge HI tails and bridges have been observed around NGC~4631 (labeled 1 to 5 in Fig. 4 of \citealt{Rand1994}). The interaction of NGC~4631 with NGC~4656 and NGC~4627 was simulated many years ago by \cite{Combes1978} indicating that the material in features 1 and 4 comes from NGC~4631, whereas that of feature 2 in the southeast and feature 3 in the northwest originate from the now dwarf elliptical galaxy NGC~4627. If so, material may also stream onto NGC~4631. This may explain the different scale heights in the different strips in NGC~4631 (see Sect.~\ref{scaleheights}) which are indeed smallest in the southeast and northwest (s1 and n3 in Table \ref{scale}). An optical CCD study of the vertical structure in NGC~4631 also showed an east-west asymmetry of the thick disk of stellar emission in the northern half \citep{verticalstructure}. Opposite the radio emission they detected the highest values of more than 2~kpc in the northwestern part of the halo. They interpret this as diffuse stellar emission and speculate that tidal debris of the satellite galaxies accreting to NGC~4631 contribute to this diffuse stellar emission. This fits with the simulations by \cite{Combes1978} mentioned above. Furthermore, \cite{Rand1993} detected two supershells in HI of 2-3~kpc in diameter of which the largest has been modeled to be formed by an HVC impact \citep{Richard}. Even extraplanar cold dust has been found at distances up to about 10~kpc in NGC~4631 which may be related to this HI supershell \citep{Neininger1999}.

As mentioned in Sect.~\ref{scaleheights} the radio halo in NGC~4631 does not show the characteristic dumbbell structure which is typical of dominating synchrotron losses. Strong star formation in the disk alone cannot be responsible for the deviating vertical scale heights in NGC~4631, as other galaxies with even higher SFR or starbursts, like NGC~253, show the usual values published by \cite{Krause2011}. All this and the fact that the scale heights are significantly larger than usual, supports the view that convection may be strong in NGC~4631. \cite{RandKulkarniHester1992} draw the same conclusion from their H$\alpha$ observations. The H$\alpha$ emission is very irregular and disturbed with a patchy high-z structure. They could not detect a smooth, extended, diffuse H$\alpha$ halo and suspected that convection is strong and matter is transported to large distances in the halo which is then responsible for the X-ray emission. 

Tidal interaction or minor mergers may also induce star formation in the disk \citep {Arshakian2009}. In fact, another galaxy with strong tidal interaction is the starburst Galaxy M~82, which had already been compared with NGC~4631 several times in the literature. Recently, \cite{Adebahr2013} was able to determine scale heights at four different wavelengths for this galaxy. All scale heights were smaller than the usual values and differed systematically between the northern and southern halves of M~82. The lower values in the south of M~82 where the IGM medium is expected to be denser than in the north because of the interaction, may lead to higher cosmic ray losses and hence smaller scale heights there. Hence tidal interaction may influence the vertical scale heights either directly by dragging out material and magnetic field from the disk into the halo or indirectly by inducing a higher star formation in the disk and/or modifying the loss processes of the cosmic rays in the halo as a result of intergalactic material produced by the tidal interaction.

\section{Conclusions}
\label{Conclusions}

We merged radio data from the Effelsberg and the VLA telescopes at $\lambda$~3.6 and 6.2~cm wavelengths of the edge-on spiral galaxy NGC~4631. The merging was done in the Fourier or u,v-plane to recover the missing spacings caused by the sampling of the VLA antennae. This is the first time maps at these frequencies of this galaxy have been obtained by combining single-dish and interferometer data. At both frequencies, we recovered more than 20\% of the total flux density by means of this combination. We analyzed the radio emission of NGC~4631 and presented the first rotation measure map between $\lambda$~3.6 and 6.2~cm. Our main conclusions are the following:

\begin{itemize}
\item We fit a single power law with the total flux-densities of NGC~4631 at four different frequencies and obtain a total integrated spectral index of $\alpha_{tot} = -0.78 \pm 0.04$. By subtracting the thermal fraction of the emission from the total flux-densities, we obtain a nonthermal integrated spectral index of $\alpha_{nth} = -0.87 \pm 0.03$.
\item The vertical scale heights in NGC~4631 vary significantly in different regions within the galaxy. Their mean values at 4.85~GHz are with 2.3~kpc (370~pc) for the thick (thin) disk higher than the mean values found in six other edge-on spiral galaxies studied so far \citep{Krause2011}. There are several indications that convection may be strong in NGC~4631, which may originally be related to its tidal interaction with its neighbouring galaxies. Because of the tidal interaction, material may even stream onto NGC~4631. Hence, this interaction may be responsible for the large spread of the values of the vertical scale heights within NGC~4631 and its larger mean values compared to those of the other six edge-on galaxies which are not strongly tidally interacting. A similar result has recently been found for M~82 \citep{Adebahr2013}.
\item The rotation measures between $\lambda$~3.6 and 6.2~cm 
are characterized by a smooth large-scale distribution, indicating that NGC~4631 has a regular magnetic field. However, no symmetric pattern with respect to the major and minor axis could be identified.
\item We detected in NGC~4631 a plane-parallel magnetic field structure along its entire disk, similar to all the other edge-on galaxies observed so far. This plane-parallel magnetic field in the disk is the expected edge-on projection of the spiral magnetic field structure observed in face-on galaxies.
\item We estimated a total magnetic field strength in the disk of NGC~4631 of $B_{t}\approx9\pm2~\rm{\mu G}$ and an ordered field strength of $B_{ord}\approx2\pm1~\rm{\mu G}$. The total magnetic field strength in the halo is comparable to that in the disk. However, the degree of uniformity in the halo is generally higher than in the disk, especially along the \emph{Northeastern extension}.
\end{itemize}

Future research on NGC~4631 will include more sensitive observations with higher angular resolution in a wider wavelength range taken with the Extended Very Large Array (EVLA or Karl G. Jansky VLA). This will allow a point by point spectral index analysis and a quantitative analysis of cosmic ray loss processes in NGC~4631. In addition, rotation measure synthesis \citep{ALLNOISES} will be applied to these multi-channel polarization observations to separate the contributions from the disk and halo in this galaxy.

\acknowledgements

We thank Krzysztof Chyzy and Michael Dumke for their help with the data reduction. We appreciate the helpful comments from Rainer Beck and the anonymous referee.

\bibliographystyle{aa}
\footnotesize
\bibliography{BIBLIOGRAFIA.bib} 
	
\end{document}